\documentclass[10pt]{llncs}
\usepackage[letterpaper,hmargin=1.2in,vmargin=1.2in]{geometry}

\usepackage{amssymb}
\usepackage{amsmath}
\usepackage[dvipdf]{graphicx}
\usepackage{url}
\usepackage[retainorgcmds]{IEEEtrantools}

\newcommand{\signed}%
    {{\unskip\nobreak\hfill\penalty50
      \hskip2em\hbox{}\nobreak\hfil $\blacksquare$
      \parfillskip=0pt \finalhyphendemerits=0 \par}}

\newcommand{\shorten}[1]{}

\begin{document}
\pagestyle{empty}
\title{The Digital Signature Scheme MQQ-SIG \\ {\small Intellectual Property Statement and Technical Description}\\ {\small 10 October 2010}}
\author{
Danilo Gligoroski\inst{1} and Svein Johan Knapskog\inst{2} and Smile Markovski\inst{3} and Rune Steinsmo \O deg\aa rd\inst{2} and Rune Erlend Jensen\inst{2} and Ludovic Perret\inst{4} and Jean-Charles Faug\`{e}re\inst{5}}
%
%
%
%
%
%
%
%
\institute{Department of Telematics, Faculty of Information
Technology, Mathematics and Electrical Engineering, The Norwegian
University of Science and Technology (NTNU), O.S.Bragstads plass 2E,
N-7491 Trondheim, NORWAY, \email{danilog@item.ntnu.no}
%
%
\and Norwegian University of Science and Technology Centre for
Quantifiable Quality of Service in Communication Systems. O.S.
Bragstads plass 2E, N-7491 Trondheim, NORWAY,
\email{knapskog@Q2S.ntnu.no, rune.odegard@q2s.ntnu.no,
runeerle@stud.ntnu.no}
\and ``Ss Cyril and Methodius'' University,  Faculty of Natural
Sciences and Mathematics, Institute of Informatics, P.O.Box 162,
1000 Skopje, MACEDONIA, \email{smile@ii.edu.mk}
\and Pierre and Marie Curie University - Paris, Laboratory of
Computer Sciences, Paris 6, 104 avenue du Pr\'{e}sident Kennedy
75016 Paris – FRANCE, \email{ludovic.perret@lip6.fr}
\and UPMC, Universit\'{e} Paris 06, LIP6 INRIA, Centre
Paris-Rocquencourt, SALSA Project-team CNRS, UMR 7606, LIP6 4, place
Jussieu 75252 Paris, Cedex 5, FRANCE
\email{jean-charles.faugere@inria.fr}
 }

\maketitle

{\bfseries Abstract:} This document contains the Intellectual
Property Statement and the technical description of the MQQ-SIG - a
new public key digital signature scheme. The complete scientific
publication covering the design rationale and the security analysis
will be given in a separate publication. MQQ-SIG consists of $n -
 \frac{n}{4}$ quadratic polynomials with $n$ Boolean
variables where $n=160$, $196$, $224$ or $256$.

\vspace{0.15cm} {\bfseries Keywords:} Public Key Cryptosystems, Fast signature generation, Multivariate Quadratic Polynomials, Quasigroup String Transformations, Multivariate Quadratic Quasigroup

\section{Intellectual Property Statement}
\vspace{-0.15cm}\hspace{0.25cm}We, the seven names given in the
title of this document and undersigned on this statement, the
authors and designers of MQQ-SIG digital signature scheme, do hereby
agree to grant any interested party an irrevocable, royalty free
licence to practice, implement and use MQQ-SIG digital signature
scheme, provided our roles as authors and designers of the MQQ-SIG
digital signature scheme are recognized by the interested party as
authors and designers of the MQQ-SIG digital signature scheme.

\vspace{0.35cm}

\texttt{\ \ \ Name\ \ \ \ \ \ \ \ \ \ \ \ \ \ \ \ \ \ \ \ \ \ Signature\ \ \ \ \ \ \ \ \ \ \ Place\ \ \ \ \ \ \ \ \ \ \ \ Date}

\vspace{0.35cm}
\texttt{1. Danilo Gligoroski\ \ \ \ \ \_\_\_\_\_\_\_\_\_\_\_\_\_\_\_\_\_\_\_\_\_\_\_\_\_\_\_\ \ \ \ Trondheim \ \ \ \ \ \ \_\_\_\_\_\_\_\_\_\_\_}

\vspace{0.5cm}
\texttt{2. Svein Johan Knapskog\ \ \_\_\_\_\_\_\_\_\_\_\_\_\_\_\_\_\_\_\_\_\_\_\_\_\_\_\_\ \ \ \ Trondheim \ \ \ \ \ \ \_\_\_\_\_\_\_\_\_\_\_}

\vspace{0.5cm}
\texttt{3. Smile Markovski\ \ \ \ \ \ \ \_\_\_\_\_\_\_\_\_\_\_\_\_\_\_\_\_\_\_\_\_\_\_\_\_\_\_\ \ \ \ Skopje\ \ \ \ \ \ \ \ \ \ \_\_\_\_\_\_\_\_\_\_\_}

\vspace{0.5cm}
\texttt{4. Rune Steinsmo \O deg\aa rd \_\_\_\_\_\_\_\_\_\_\_\_\_\_\_\_\_\_\_\_\_\_\_\_\_\_\_\ \ \ \ Trondheim \ \ \ \ \ \ \_\_\_\_\_\_\_\_\_\_\_}

\vspace{0.5cm}
\texttt{5. Rune Erlend Jensen\ \ \ \ \_\_\_\_\_\_\_\_\_\_\_\_\_\_\_\_\_\_\_\_\_\_\_\_\_\_\_\ \ \ \ Trondheim \ \ \ \ \ \ \_\_\_\_\_\_\_\_\_\_\_}

\vspace{0.5cm}
\texttt{6. Ludovic Perret\ \ \ \ \ \ \ \ \_\_\_\_\_\_\_\_\_\_\_\_\_\_\_\_\_\_\_\_\_\_\_\_\_\_\_\ \ \ \ Paris\ \ \ \ \ \ \ \ \ \ \ \_\_\_\_\_\_\_\_\_\_\_}

\vspace{0.5cm}
\texttt{7. Jean-Charles Faug\`{e}re\ \ \_\_\_\_\_\_\_\_\_\_\_\_\_\_\_\_\_\_\_\_\_\_\_\_\_\_\_\ \ \ \ Paris\ \ \ \ \ \ \ \ \ \ \ \_\_\_\_\_\_\_\_\_\_\_}

\section{Description of the MQQ-SIG digital signature scheme}\label{SecDescr}
A generic description for our scheme can be expressed as a
$\frac{3}{4}$ truncation of a typical multivariate quadratic system:
$\mathbf{S} \circ P' \circ \mathbf{S'}:\{0,1\}^n \to \{0,1\}^n$
where $\mathbf{S'} = \mathbf{S} \cdot \mathbf{x} + \mathbf{v}$ (i.e.
$\mathbf{S'}$ is a bijective affine transformation), $\mathbf{S}$ is
a nonsingular linear transformation, and $P'$ is a bijective
multivariate quadratic mapping on $\{0,1\}^n$.

The bijective multivariate quadratic mapping $P':\{0,1\}^n
\rightarrow \{0,1\}^n$ is defined in Table \ref{NonlinearPart}.
\vspace{-0.35cm}\begin{table}[!h]
  \centering
  {\scriptsize
\begin{tabular}{lr}
\begin{tabular}{|c|}
  \hline
   \parbox{12.0cm}{\ \\Bijective multivariate quadratic mapping  {$P'(\mathbf{x})$}\\ \ } \\ %
  \hline
  \parbox{12.0cm}{\ \\ {\bf Input:} A vector $\mathbf{x}=(f_1,\ldots,f_n)$ of $n$ linear Boolean functions of $n$ variables. We implicitly suppose that a multivariate quadratic quasigroup $*$ is previously defined, and that $n=32k$, $k\in \{5, 6, 7, 8\}$ is also previously determined.\\ \ }\\
  \hline
  \parbox{12.0cm}{{\bf Output:} $8$ linear expressions $P'_i(x_1,\ldots,x_n), i=1,\ldots,8$ and $n-8$ multivariate quadratic polynomials $P'_i(x_1,\ldots,x_n), i=9,\ldots,n$\\ \ }\\
  \hline
  \begin{tabular}{l}
  \parbox{12.0cm}{\ \\1. Represent a vector $\mathbf{x}=(f_1,\ldots,f_n)$ of $n$ linear Boolean functions of $n$ variables $x_1,\dots, x_n$, as a string $\mathbf{x}=X_1 \ldots X_{\frac{n}{8}}$ where $X_i$ are vectors of dimension 8;}\\
  \parbox{12.0cm}{2. Compute $\mathbf{y}=Y_1 \ldots Y_{\frac{n}{8}}$ where:  $Y_1 = X_1$, $Y_{j+1} = X_{j} * X_{j+1}$, for even $j=2, 4, \ldots$, and $Y_{j+1} = X_{j+1} * X_{j}$, for odd $j=3, 5, \ldots$}\\
  \parbox{12.0cm}{3. Output: $\mathbf{y}$.}
 \end{tabular}\\
 \hline
  \end{tabular}
\end{tabular}
}
  \caption{Definition of the bijective multivariate quadratic mapping $P':\{0,1\}^n \rightarrow
\{0,1\}^n$}\label{NonlinearPart}
\end{table}

\vspace{-0.35cm} The algorithm for generating the public and private
key is defined in the Table \ref{GeneratingKeys}.
\vspace{-0.35cm}\begin{table}[!h]
  \centering
{\scriptsize
\begin{tabular}{lr}
\begin{tabular}{|c|}
  \hline
   \parbox{12.0cm}{\ \\Algorithm for generating Public and Private key for the MQQ-SIG scheme\\ \ }\\ %
  \hline
  \parbox{12.0cm}{\ \\ {\bf Input:} Integer $n$, where $n=32\times k$ and $k\in\{5, 6, 7, 8\}$.\\ \ }\\
  \hline
  \parbox{12.0cm}{\ \\ {\bf Output:} Public key $\mathbf{P}$: $n -  \frac{n}{4}$ multivariate quadratic polynomials  $P_i(x_1,\ldots,x_n),\
  i=1+ \frac{n}{4},\ldots, n$, Private key: Two permutations $\sigma_1$ and $\sigma_K$ of the numbers $\{1,\ldots,n\}$, and 81 bytes for encoding a quasigroup $*$ .\\ \ }\\
  \hline
  \begin{tabular}{l}
  \parbox{12.0cm}{\ \\ 1. Generate an MQQ $*$ according to equations (\ref{BilinearMQQ}) \dots (\ref{RankMatrices}).}\\
  \parbox{12.0cm}{2. Generate a nonsingular ${n \times n}$ Boolean matrix $\mathbf{S}$ and affine transformation $\mathbf{S'}$ according to equations (\ref{SInv}), \dots, (\ref{VectorV}).}\\
  \parbox{12.0cm}{3. Compute $\mathbf{y}=\mathbf{S}(P'(\mathbf{S'}(\mathbf{x})))$, where $\mathbf{x}=(x_1,\ldots,x_n)$.}\\
  \parbox{12.0cm}{4. Output: The public key is $\mathbf{y}$ as $n - \frac{n}{4}$ multivariate quadratic
  polynomials $P_i(x_1,\ldots,x_n)\ i=1+ \frac{n}{4},\ldots, n $, and the private key is the tuple
  $(\sigma_1,\sigma_K,*)$.\\ \ }
 \end{tabular}\\
 \hline
  \end{tabular}
\end{tabular}
 }
  \caption{Generating the public and private key}\label{GeneratingKeys}
\end{table}

\vspace{-0.25cm}The algorithm for signing by the private key
$(\sigma_1,\sigma_K,*)$ is defined in Table \ref{DecryptionSigning}.
\begin{table}[!h]
  \centering
{\scriptsize
\begin{tabular}{lr}
\begin{tabular}{|c|}
  \hline
   \parbox{12.0cm}{\ \\Algorithm for digital signature with the private key $(\sigma_1,\sigma_K,*)$\\ \ }\\ %
  \hline
  \parbox{12.0cm}{\ \\{\bf Input:} A document $M$ to be signed.\\ \ }\\
  \hline
  \parbox{12.0cm}{\ \\{\bf Output:} A signature $\mathbf{sig}=(x_1,\ldots,x_{n})$.\\ \ }\\
  \hline
  \begin{tabular}{l}
  \parbox{12.0cm}{\ \\1. Compute $\mathbf{y}=(y_1,\ldots,y_n) = H(M)|_n$, where $M$ is the message to be signed,
  $H()$ is a standardized cryptographic hash function such as SHA-1, or SHA-2,
  with a hash output of not less than $n$ bits. The notation $H(M)|_n$ denotes
  the least significant $n$ bits from the hash output $H(M)$.}\\
  \parbox{12.0cm}{2. Set $\mathbf{y'}=\mathbf{S}^{-1}(\mathbf{y})$.}\\
  \parbox{12.0cm}{3. Represent $\mathbf{y'}$ as $\mathbf{y'}=Y_1 \ldots Y_\frac{n}{8}$ where $Y_i$ are Boolean vectors of dimension 8.}\\
  \parbox{12.0cm}{4. By using the left and right parastrophes $\setminus$ and $/$ of the quasigroup $*$
  compute $\mathbf{x'}=X_1 \ldots X_\frac{n}{8}$, such that: $X_1 = Y_1$, $X_j = X_{j-1} \setminus Y_{j}$, for even $j=2, 4, \ldots$, and
  $X_j = Y_{j} / X_{j-1}$, for odd $j=3, 5, \ldots$.}\\
  \parbox{12.0cm}{5. Compute $\mathbf{x}=\mathbf{S}^{-1}(\mathbf{x'}) + \mathbf{v} = (x_1,\ldots,x_n)$.}\\
  \parbox{12.0cm}{6. The MQQ-SIG digital signature of the document $M$ is the vector $\mathbf{sig}=(x_1,\ldots,x_n)$.}\\
 \end{tabular}\\
 \hline
  \end{tabular}
\end{tabular}
 }
  \caption{Digital signing}\label{DecryptionSigning}
\end{table}

\vspace{-0.25cm}The algorithm for signature verification with the
public key $\mathbf{P}=\{ P_i(x_1,\ldots,x_n)\ |\ i=1+
\frac{n}{4},\ldots,n \}$ is given in Table \ref{Verification}.
\begin{table}[!h]
  \centering
{\scriptsize
\begin{tabular}{lr}
\begin{tabular}{|c|}
  \hline
   \parbox{12.0cm}{\ \\Algorithm for signature verification with a public key $\mathbf{P}=\{ P_i(x_1,\ldots,x_n)\ |\ i=1+  \frac{n}{4},\ldots,{n }\}$\\ \ }\\ %
  \hline
  \parbox{12.0cm}{{\bf Input:} A document $M$ and its signature $\mathbf{sig}=(x_1,\ldots,x_{n})$.}\\
  \hline
  \parbox{12.0cm}{{\bf Output:} TRUE or FALSE.}\\
  \hline
  \begin{tabular}{l}
  \parbox{12.0cm}{\ \\1. Compute $\mathbf{y}=(y_{1+ \frac{n}{4}},\ldots,y_n) = H(M)|_{n- \frac{n}{4}}$, where $M$ is the signed message,
  $H()$ is a standardized cryptographic hash function such as SHA-1, or SHA-2,
  with a hash output of not less than $n$ bits, and the notation $H(M)|_{n- \frac{n}{4}}$ denotes
  the least significant $n- \frac{n}{4}$ bits from the hash output $H(M)$.}\\
  \parbox{12.0cm}{2. Compute $\mathbf{z}=(z_{1+  \frac{n}{4}},\ldots,z_{n})=\mathbf{P}(\mathbf{sig})$.}\\
  \parbox{12.0cm}{3. If $\mathbf{z} = \mathbf{y}$ then return TRUE, else return FALSE.}\\
 \end{tabular}\\
 \hline
  \end{tabular}
\end{tabular}
 }
  \caption{Digital verification}\label{Verification}
\end{table}

\section{Multivariate Quadratic Quasigroups}\label{SecPrelim}
A Multivariate Quadratic Quasigroup (MQQ) $*$ of order $2^d$ used in
this version of MQQ-SIG can be described shortly by the following
expression:
\begin{equation}\label{BilinearMQQ}
\mathbf{x} * \mathbf{y} \equiv \mathbf{B} \cdot
\mathbf{U}(\mathbf{x}) \cdot \mathbf{A_2} \cdot \mathbf{y} +
\mathbf{B} \cdot \mathbf{A_1} \cdot \mathbf{x} + \mathbf{c}
\end{equation}
where $\mathbf{x}=(x_1,\dots,x_d)$, $\mathbf{y}=(y_1,\dots,y_d)$,
the matrices $\mathbf{A_1}$, $\mathbf{A_2}$ and $\mathbf{B}$ are
nonsingular in $GF(2)$, of size $d\times d$, the vector $\mathbf{c}$
is a random $d$-dimensional vector with elements in $GF(2)$ and all
of them are generated by a uniformly random process. The matrix
$\mathbf{U}(\mathbf{x})$ is an upper triangular matrix with all
diagonal elements equal to 1, and the elements above the main
diagonal are linear expressions of the variables of
$\mathbf{x}=(x_1,\dots,x_d)$. It is computed by the following
expression:
\begin{equation}\label{Ux}
\mathbf{U}(\mathbf{x}) = I + \sum_{i=1}^{d-1} \mathbf{U}_i \cdot
\mathbf{A_1} \cdot \mathbf{x},
\end{equation}
where the matrices $\mathbf{U}_i$ have all elements 0 except the
elements in the rows from $\{1,\ldots,i\}$ that are strictly above
the main diagonal. Those elements can be either 0 or 1.

Once we have a multivariate quadratic quasigroup
$$*_{vv}(x_1,\dots,x_d,y_1,\dots,y_d)=(f_1(x_1,\dots,x_d,y_1,\dots,y_d),...,f_d(x_1,\dots,x_d,y_1,\dots,y_d))$$
we will be interested in those quasigroups that will satisfy the
following conditions:
\begin{IEEEeqnarray}{rCl}
\label{Ranks}
\forall i \in \{1, \ldots, d\}, Rank(\mathbf{B}_{f_i}) & \geq & 2 d - 4, \IEEEyessubnumber\\
\exists j \in \{1, \ldots, d\}, && Rank(\mathbf{B}_{f_j}) = 2 d - 2
\IEEEyessubnumber\
\end{IEEEeqnarray}
where matrices $\mathbf{B}_{f_i}$ are $2d \times 2d$ Boolean
matrices defined from the expressions $f_i$ as
\begin{equation}\label{RankMatrices}
\mathbf{B}_{f_i}=[b_{j,k}], \ b_{j,d+k}=b_{d+k,j}=1,\ \mbox{iff}\
x_j y_k \mbox{ is a term in } f_i.
\end{equation}

\begin{proposition}
For $d=8$, a multivariate quadratic quasigroup that satisfies the
conditions (\ref{BilinearMQQ}), \dots, (\ref{RankMatrices}) can be
encoded in a unique way with 81 bytes.
\end{proposition}

\section{Nonsingular Boolean matrices in MQQ-SIG}
In MQQ-SIG the nonsingular matrices $\mathbf{S}$ are defined by the
following expression:
\begin{equation}\label{SInv}
\mathbf{S}^{-1} = \sum_{i=1}^K I_{\sigma_i},
\end{equation}
where $I_{\sigma_i}, \ i=\{1, 2, \ldots, K\}$ are permutation
matrices of size $n=32\times k$ and where permutations $\sigma_i$
are permutations on $n$ elements. They are defined by the following
expressions:
\begin{equation}\label{K}
K = \left\{
\begin{array}{cll}
k & , & \mbox{if\ } k \mbox{\ is odd,}\\
k+1 & , & \mbox{if\ } k \mbox{\ is even}\\
\end{array} \right.
\end{equation}
\begin{equation}\label{permutations}
\left\{
\begin{array}{cll}
\sigma_1 & - & \mbox{random permutation on }
\{1, 2, \ldots n\} \mbox{ satisfying the condition (\ref{First8})},\\
\sigma_2 & = & RotateLeft(\sigma_1,32) \mbox{ satisfying the condition (\ref{First8})},\\
\sigma_3 & = & RotateLeft(\sigma_2,64) \mbox{ satisfying the condition (\ref{First8})},\\
\sigma_j & = & RotateLeft(\sigma_{j-1},32), \mbox{\ for\ } j=4, \ldots, K-1, \mbox{ satisfying the condition (\ref{First8})},\\
\sigma_K & - & \mbox{random permutation on }
\{1, 2, \ldots n\} \mbox{ satisfying the condition (\ref{First8})}\\
\end{array}
\right.
\end{equation}
\begin{equation}\label{First8}
\sigma_\nu = \left(
\begin{array}{cccccccc}
1   & 2    & \ldots &   8 &   9 & \ldots  &  n-1 & n \\
s^{(\nu)}_1 & s^{(\nu)}_ 2 & \ldots & s^{(\nu)}_8 & s^{(\nu)}_9 &
\ldots & s^{(\nu)}_{n-1} & s^{(\nu)}_n
\end{array}
 \right), \ \{s^{(\nu)}_1, s^{(\nu)}_ 2, \ldots, s^{(\nu)}_8\} \bigcap \{1, 2, \ldots, 8\} =
 \emptyset
\end{equation}
where $RotateLeft(\sigma, l)$ denotes a permutation obtained from
the permutation $\sigma$ by rotating it to the left for $l$
positions.

We require an additional condition to be fulfilled by the
permutations $\sigma_1, \dots, \sigma_K$:
\begin{equation}\label{LatinRectangle}
L = \left[
\begin{array}{ccccccc}
\sigma_1 \\
\sigma_2 \\
\vdots\\
\sigma_{K-1} \\
\sigma_K \\
\end{array}
 \right], \ \mbox{is a Latin Rectangle.}
\end{equation}

Once we have a nonsingular matrix $\mathbf{S}^{-1}$ we will compute
its inverse obtaining $$\mathbf{S} = (\mathbf{S}^{-1})^{-1}$$ and
from there we will obtain the affine transformation
\begin{equation}\label{SPrim}
\mathbf{S'}(\mathbf{x}) = \mathbf{S} \cdot \mathbf{x} + \mathbf{v},
\end{equation}
where the vector $\mathbf{v}$ is $n$--dimensional Boolean vector
defined from the values of the permutation $\sigma_K$ by the
following expression:
\begin{equation}\label{VectorV}
\mathbf{v} = (v_1, v_2, \ldots, v_n), \ \mbox{where } v_i = \left(
\frac{s^{(K)}_{64+\lceil \frac{i}{4} \rceil}}{2^{i \mbox{ mod }4}}
\right) \mbox{ mod }2.
\end{equation}

In words: we construct the bits of the vector $\mathbf{v}$ by taking
the four least significant bits of the values $s^{(K)}_{65},$
$\ldots,$ $s^{(K)}_{64+\frac{n}{4}}$ in the permutation $\sigma_K$.

\begin{proposition}
The linear transformation $\mathbf{S}^{-1}$ can be encoded in a
unique way with $2n$ bytes.
\end{proposition}

\newpage
\section{Characteristics of the MQQ-SIG digital signature scheme}

The main characteristics of our MQQ-SIG digital signature scheme can
be briefly summarized as follows:

    $\quad\bullet$\ there is no message expansion;

    $\quad\bullet$\ the length of the signature is $n$ bits where ($n=160, 192, 224$ or $256$);

    $\quad\bullet$\ its conjectured security level is $2^{\frac{n}{2}}$;

    $\quad\bullet$\ its verification speed is comparable to the speed of other
multivariate quadratic PKCs;

    $\quad\bullet$\ in software its signing speed is in the range of 500--5,000 times
faster than RSA and ECC schemes;

    $\quad\bullet$\ in hardware its signing or verification speed is more than 10,000 times
faster than RSA and ECC schemes;

    $\quad\bullet$\ it is also well suited for producing short signatures in smart cards
    and RFIDs;
%
%
%
\subsection{The size of the public and the private key}
The size of the public key is $0.75 \times
n\times(1+\frac{n(n+1)}{2})$ bits. The private key of our scheme is
the tuple $(\sigma_1,\sigma_K,*)$. The corresponding memory size
needed for storage of the private key is $2n + 81$ bytes. In Table
\ref{PublicKeyMemory} we give the size of the public key (in KBytes)
and the size of the private key (in bytes) for $n\in\{160,192,224,
256\}$.
\begin{table}[h!]
\begin{center}
{\footnotesize
\begin{tabular}{c|ccc@{.}l|cccccc|}
$n$ & \multicolumn{4}{c|}{$\begin{array}{c}\mbox{Size of the}\\ \mbox{public key (KBytes)}\end{array}$ }& \multicolumn{5}{c|}{$\begin{array}{c}\mbox{Size of the}\\ \mbox{private key (bytes)}\end{array}$ }\\
\hline
160 & && 188&69 & &&&& 401 \\
192 & && 325&71 & &&&& 465 \\
224 & && 516&82 & &&&& 529 \\
256 & && 771&02 & &&&& 593 \\
\end{tabular}\bigskip
}
\end{center}
\vspace{-0.5cm} \caption{Memory size in KBytes for the public key
and in bytes for the private key} \label{PublicKeyMemory}
\end{table}

\shorten{

}

\end{document}